\AtBeginDocument{%
  }
    
\documentclass[sigconf, 9pt]{acmart}

\usepackage{amsmath,amssymb,amsfonts}
\usepackage{algpseudocode}
\usepackage{algorithm}
\usepackage{graphicx}
\usepackage{textcomp}
\usepackage{xcolor}
\usepackage{hyperref}
\usepackage{soul}
\usepackage{physics,array}
\usepackage{multirow}
\usepackage{tikz}
\usepackage{adjustbox}
\usepackage{geometry}
\usepackage{changepage}
\usepackage{pifont}
\usepackage{amsmath}
\usepackage{amssymb}
\setcopyright{acmlicensed}
\copyrightyear{2018}
\acmYear{2018}
\acmDOI{XXXXXXX.XXXXXXX}

\renewcommand\footnotetextcopyrightpermission[1]{}

\acmBooktitle{Woodstock '18: ACM Symposium on Neural Gaze Detection,
 June 03--05, 2018, Woodstock, NY}
\acmISBN{978-1-4503-XXXX-X/18/06}

\begin{document}
\fancyhead{}



\title{The Art of Optimizing T-Depth for Quantum Error Correction in Large-Scale Quantum Computing}
    

\author{Avimita Chatterjee}
\orcid{1234-5678-9012}
\affiliation{%
  \institution{Pennsylvania State University}
  \city{State College}
  \state{PA}
  \country{USA}
}
\email{amc8313@psu.edu}

\author{Archisman Ghosh}
\affiliation{%
  \institution{Pennsylvania State University}
  \city{State College}
  \state{PA}
  \country{USA}
  }
\email{apg6127@psu.edu}

\author{Swaroop Ghosh}
\affiliation{%
  \institution{Pennsylvania State University}
  \city{State College}
  \state{PA}
  \country{USA}
}
\email{szg212@psu.edu}


\begin{abstract}

Quantum Error Correction (QEC), combined with magic state distillation, ensures fault tolerance in large-scale quantum computation. To apply QEC, a circuit must first be transformed into a non-Clifford (or T) gate set. T-depth, the number of sequential T-gate layers, determines the magic state cost, impacting both spatial and temporal overhead. Minimizing T-depth is crucial for optimizing resource efficiency in fault-tolerant quantum computing. While QEC scalability has been widely studied, T-depth reduction remains an overlooked challenge. We establish that T-depth reduction is an NP-hard problem and systematically evaluate multiple approximation techniques: greedy, divide-and-conquer, Lookahead-based brute force, and graph-based. The Lookahead-based brute-force algorithm (partition size 4) performs best, optimizing 90\% of reducible cases (i.e., circuits where at least one algorithm achieved optimization) with an average T-depth reduction of around 51\%. Additionally, we introduce an expansion factor-based identity gate insertion strategy, leveraging controlled redundancy to achieve deeper reductions in circuits initially classified as non-reducible. With this approach, we successfully convert up to 25\% of non-reducible circuits into reducible ones, while achieving an additional average reduction of up to 11.8\%. Furthermore, we analyze the impact of different expansion factor values and explore how varying the partition size in the Lookahead-based brute-force algorithm influences the quality of T-depth reduction.
\end{abstract}

\maketitle

\section{Introduction} \label{sec:introduction}

Quantum error correction (QEC)~\cite{preskill1998reliable, lidar2013quantum, roffe2019quantum} is fundamental to quantum computing, mitigating the effects of noise and decoherence~\cite{clerk2010introduction} by exponentially suppressing errors. This suppression is crucial for reducing error rates to levels necessary for practical quantum computation. Among various QEC strategies, the surface code~\cite{fowler2012surface, kitaev2003fault} is one of the most extensively studied and implemented due to its reliance on local interactions, a high fault-tolerance threshold of approximately 0.7\%~\cite{google2023suppressing}, and its capability to enable universal quantum computation when integrated with a magic state factory~\cite{bravyi2012magic, litinski2019magic}.

A well-established approach to low-overhead fault-tolerant quantum computation is based on the Clifford + T gate formalism~\cite{kliuchnikov2012fast, brown2017poking}. In this model, Clifford gates can be implemented fault-tolerantly using surface codes, whereas T gates, which are non-Clifford, must be injected via magic state distillation~\cite{haah2018codes, bravyi2005universal}. The implementation of T gates is particularly resource-intensive. While Clifford gates can be executed directly, each T gate requires the consumption of a high-fidelity magic state, defined as \( |m\rangle = |0\rangle + e^{i\pi/4}|1\rangle \)~\cite{litinski2018quantum}. However, initially prepared magic states are noisy and require purification through magic state distillation, a costly process in terms of both qubits and time. To mitigate this overhead, the Pauli-based computation framework~\cite{gottesman1998heisenberg} restructures quantum circuits by commuting and eliminating unnecessary Clifford gates while isolating non-Clifford gate blocks for targeted error correction~\cite{litinski2019game}.

\textbf{Background:} A \(\pi/8\) gate is a unitary transformation that applies a controlled phase shift and is fundamental in quantum computation beyond the Clifford group. 
It is referred to as a \(\pi/8\) gate because, when expressed in Pauli exponential form, it is written as \( T = e^{-i\pi/8} Z \), where \( Z \) is the Pauli-Z operator. More generally, Pauli \(\pi/8\) gates are written as \( e^{-i\pi/8} P \), where \( P \in \{I, X, Y, Z\} \), representing fractional Pauli rotations. Throughout this paper, we use the notation \( \pm P \) instead of explicitly writing \( e^{\pm i\pi/8} P \). Specifically, we define \( +P \) to represent \( e^{i\pi/8} P \) and \( -P \) to represent \( e^{-i\pi/8} P \). This convention provides a compressed representation of Pauli \(\pi/8\) gates, simplifying the analysis of commutative products and T-depth reduction in quantum circuits.

When an initial quantum circuit undergoes quantum error correction (QEC), it consists of Clifford and non-Clifford (T) gates. Before this circuit can be mapped onto the surface code for processing, it must be transformed such that only T gates remain. These transformations optimize non-Clifford gates by commuting them through Clifford gates, reducing circuit complexity while maintaining computational equivalence.  
Clifford Pauli product rotations, represented as \( e^{-i\pi/4} P \), can be commuted past non-Clifford Pauli product rotations. The commutation rules governing these operations are as follows~\cite{litinski2019game}:  
If \( P'P = P'P \), meaning the operators commute, then \( P \) can pass through \( P' \) without altering the operator.  
- If \( P'P = -P'P \), meaning the operators anti-commute, then commuting \( P \) past \( P' \) introduces a phase factor \( i \), transforming \( P' \) into \( iP' \) for the non-Clifford operator.  
Additionally, measurement operations are optimized by absorbing Clifford gates into the measurement operators. This process removes all Clifford gates from the circuit, transforming it into a grid-like structure where qubits form the rows, and each column represents the non-Clifford gate acting on each qubit. A step-by-step example of how circuits are fully reduced into this form can be found in ~\cite{litinski2019game}.


Once this reduced circuit is mapped onto the surface code, magic state distillation protocols are required to implement T gates fault-tolerantly. Each T gate layer necessitates a corresponding distilled magic state. A distillation protocol follows an \(N\)-to-\(K\) scheme, where \(N\) represents the number of input magic states, and \(K\) is the number of distilled, high-fidelity states produced. However, magic state distillation is both time and space-intensive, as it involves preparing high-fidelity magic states from noisy ones~\cite{haah2018codes, bravyi2005universal, litinski2018quantum}. 
Consequently, reducing the number of T-gate columns (T-depth) directly reduces the number of magic states required, significantly reducing both qubit overhead and overall processing time.

In a quantum circuit where only non-Clifford (or T) gates remain, each column represents a separate T-gate layer, meaning the number of layers is initially equal to the number of columns. The key observation is that columns can be merged if and only if all elements of one column commute with all elements of the other column. This follows directly from the commutative product rules (Table~\ref{tab:multiplication_table}) of Pauli operators, which dictate when two T-gate columns can be combined without changing the computation. Since T gates only apply single-qubit phase rotations and do not introduce entanglement, the order in which commuting columns are merged does not affect the final quantum state. This is because commuting operations can be applied in any sequence without altering the overall transformation. Thus, rather than following a fixed order, the merging process can be performed in any sequence, provided that only commuting columns are combined at each step. 

\begin{figure}
    \centering
    \includegraphics[width=1\linewidth]{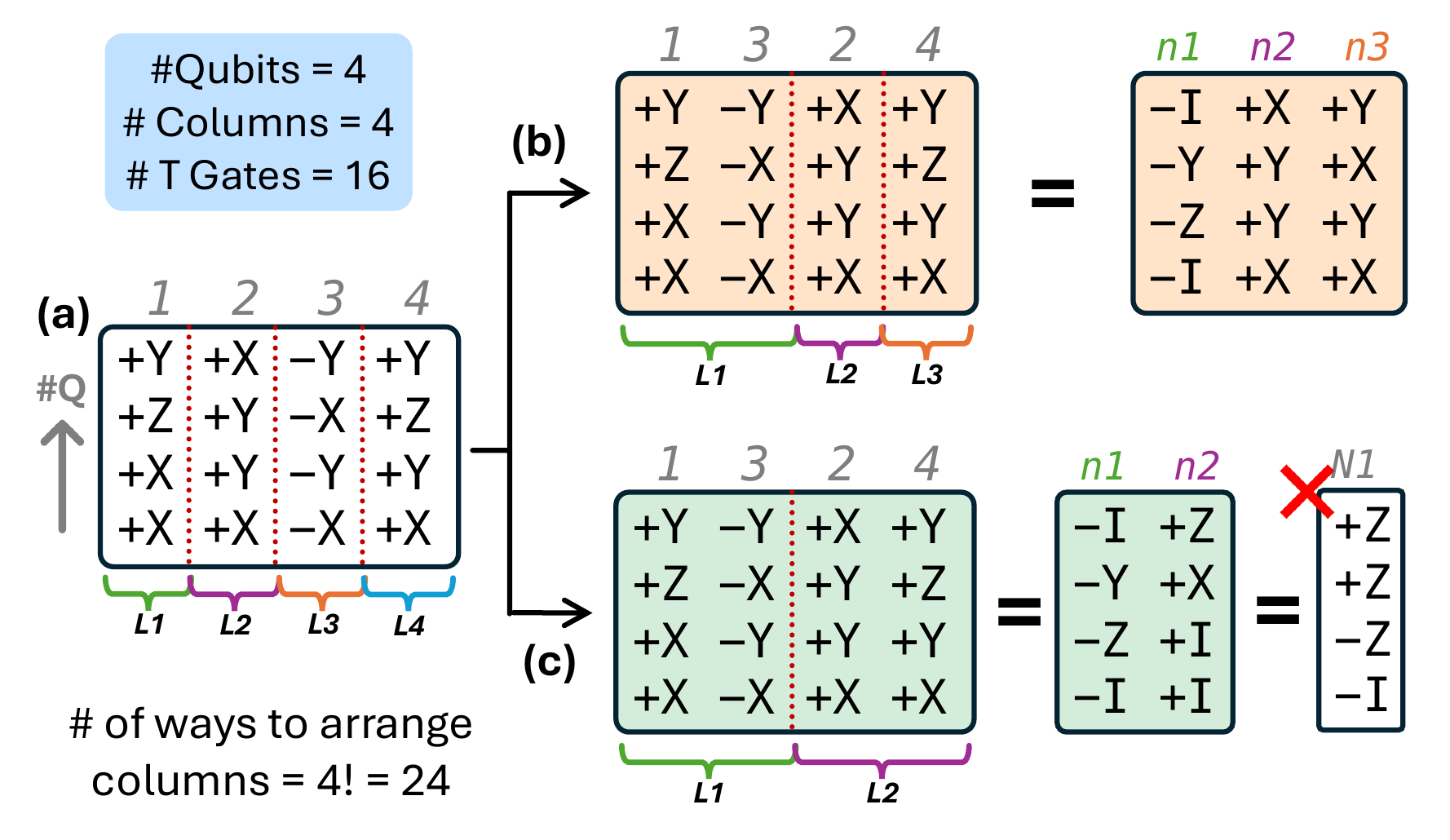}
    \caption{ \textbf{Different Layering Strategies for a Circuit:} 
   An example circuit with 4 qubits, 4 columns, and 16 T gates, demonstrating three layering approaches. (a) Preserves the original structure. (b) Reorders columns, merging 1 and 3 to reduce the depth to 3 layers. (c) Further reduces the depth to 2 layers by merging 1 with 3 and 2 with 4, following commutative product rules from Table~\ref{tab:multiplication_table}. An attempted merge of columns 1 and 2 fails due to inconsistent T gate phases, emphasizing the need for careful selection.
    }
    \label{fig:example_layer}
    \vspace{-5mm}
\end{figure}

Fig.~\ref{fig:example_layer} illustrates a 4-qubit, 4-column circuit with 16 T gates, demonstrating three distinct approaches to layer formation. With 4 columns, there are $4! = 24$ possible ways to arrange and attempt to merge them, of which we present three examples.
In the first approach (a), the circuit retains all 4 layers, preserving both the ordering and number of columns from the original circuit. The second approach (b) adopts a different column arrangement and reduces the depth to 3 layers by merging columns 1 and 3. The third approach (c) follows the same arrangement as the second but further minimizes the depth to 2 layers by merging columns 1 with 3 and 2 with 4. Additionally, the example explores an attempted merge of the newly formed columns 1 and 2, which fails due to inconsistent T gate phases within the column.
This example highlights the importance of exploring various column arrangements to determine the optimal merging strategy that minimizes the number of layers.

\begin{table}[]
\centering
\fontsize{8.5pt}{9.5pt}\selectfont
\caption{Commutative Products of Pauli Matrices}
\begin{tabular}{cc||cccc|cccc}
\hline \hline
\multicolumn{2}{c||}{\multirow{2}{*}{\textbf{A.B}}}             & \multicolumn{8}{c}{\textbf{A}}                                                                                                     \\ \cline{3-10} 
\multicolumn{2}{c||}{}                                          & \textbf{+I} & \textbf{+X} & \textbf{+Y} & \textbf{+Z} & \textbf{-I} & \textbf{-X} & \textbf{-Y} & \textbf{-Z} \\ \hline \hline
\multicolumn{1}{c|}{\multirow{8}{*}{\textbf{B}}} & \textbf{+I} & +I          & +X          & +Y          & +Z          & -I          & -X          & -Y          & -Z          \\
\multicolumn{1}{c|}{}                            & \textbf{+X} & +X          & +I          & -iZ         & +iY         & -X          & -I          & +iZ         & -iY         \\
\multicolumn{1}{c|}{}                            & \textbf{+Y} & +Y          & +iZ         & +I          & -iX         & -Y          & -iZ         & -I          & +iX         \\
\multicolumn{1}{c|}{}                            & \textbf{+Z} & +Z          & -iY         & +iX         & +I          & -Z          & +iY         & -iX         & -I          \\ \cline{2-10} 
\multicolumn{1}{c|}{}                            & \textbf{-I} & -I          & -X          & -Y          & -Z          & +I          & +X          & +Y          & +Z          \\
\multicolumn{1}{c|}{}                            & \textbf{-X} & -X          & -I          & +iZ         & -iY         & +X          & +I          & -iZ         & +iY         \\
\multicolumn{1}{c|}{}                            & \textbf{-Y} & -Y          & -iZ         & -I          & +iX         & +Y          & +iZ         & +I          & -iX         \\
\multicolumn{1}{c|}{}                            & \textbf{-Z} & -Z          & +iY         & -iX         & -I          & +Z          & -iY         & +iX         & +I          \\ \hline \hline
\end{tabular}
\label{tab:multiplication_table}
\end{table}

\textbf{Motivation:} Current research primarily focuses on scaling quantum error correction codes for large-scale quantum computers; however, an important gap remains in optimizing the circuit before mapping it to QECC, which could significantly reduce the overall resource overhead.
By systematically identifying and merging as many commuting columns as possible at each step, the circuit depth can be progressively reduced. This process needs to continue iteratively until no further merges are possible. The final number of columns after merging should represent the minimum number of T-gate layers required, ensuring the lowest possible T-depth for the circuit. When the final optimized circuit is integrated with a QECC, it will require an optimal number of distilled magic states.

Given a quantum circuit of T gates, the goal is to minimize T-depth, the number of sequential T layers constrained by commutation. A naïve approach would evaluate all $c!$ permutations of columns to determine the optimal sequence of commutative products. At each step, merging two columns requires verifying Pauli commutation relations and the impact of T-gate placement. Once the first level of reduction is completed, the process repeats at the reduced level, continuing iteratively through successive levels while examining every column permutation at each stage. This method always guarantees the optimal result, as it exhaustively explores all possible column reorderings to find the most effective configuration, ultimately producing the final reduced circuit with the minimum number of layers. Throughout this paper, we refer to this as the Brute-Force approach.
The time complexity of this approach for a circuit with $n$ qubits and $c$ columns is $O(c! \cdot nc^2)$.
This exponential scaling in the worst case indicates that the problem becomes computationally intractable. This problem can be formally proven to be NP-hard by reducing it from a known NP-hard problem, such as Boolean satisfiability~\cite{van2023optimising}. 
This implies that the problem lacks a known polynomial-time algorithm to efficiently solve all cases, necessitating the use of approximation techniques. However, no study has comprehensively examined potential approximation strategies for this specific NP-hard problem to understand their effectiveness.

\begin{figure}
    \centering
    \includegraphics[width=1\linewidth]{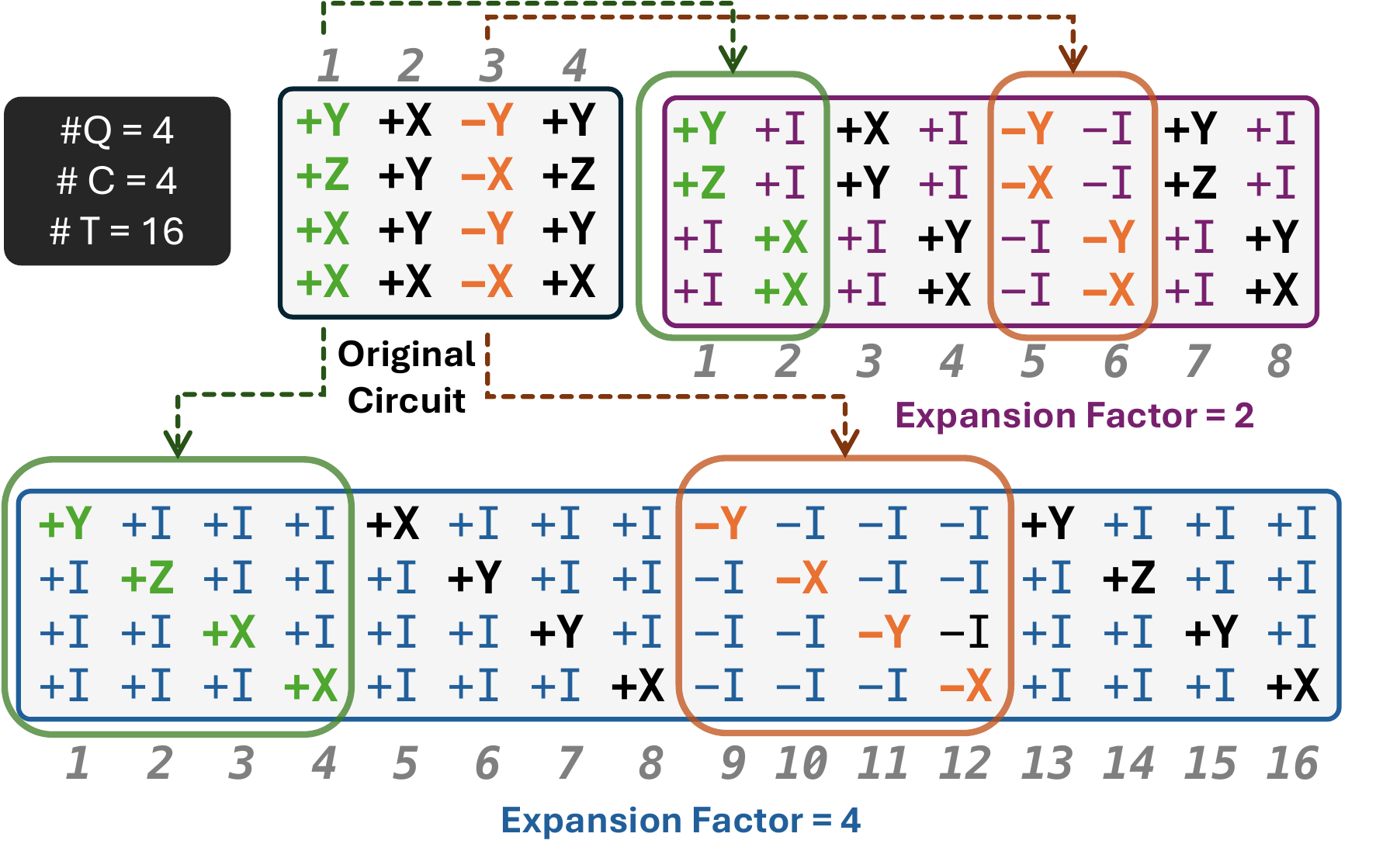}
    \caption{Illustration of Circuit Expansion with Different Expansion Factors: 
    An example of an original circuit with four columns, four qubits, and sixteen T gates, demonstrating the application of expansion factors of 2 and 4.
    }
    \label{fig:times_factor_figure}
    \vspace{-5mm}
\end{figure}

\textbf{Contributions:}
The aim of this paper is to develop and analyze approximation techniques for reducing the T-depth of quantum circuits, thereby optimizing resource efficiency in fault-tolerant quantum computing.
We develop four approximation techniques to achieve the above aim: Greedy, Divide and Conquer, Lookahead-based brute-force, and Graph-based algorithms. 
Among our diverse set of circuits, we observe that 34\% are non-reducible, meaning their column count remains unchanged regardless of the algorithm used. Among the approximation methods, the Lookahead-based brute-force algorithm with a partition size of 4 performs the best, achieving optimal results in 90\% of the reducible cases with an average column reduction of approximately 51\%.
%
%
Our analysis reveals that high T-gate density is the primary limiting factor in circuit optimization for initially non-reducible circuits. This challenge is further amplified by circuit depth, particularly when combined with a high concentration of T gates.
To address initially non-reducible circuits, we introduce an expansion factor-based technique, which strategically inserts redundant identity gates without altering the circuit’s functionality. Fig.~\ref{fig:times_factor_figure} illustrates an example of an original circuit with four columns, four qubits, and sixteen T gates, showcasing expansion factors of 2 and 4. This transformation allows us to apply the same approximation techniques to the expanded circuits, effectively enabling further reductions. Through this approach, we successfully reduce approximately 25\% of the circuits initially classified as non-reducible.
Additionally, we investigate the impact of different expansion factors on the quality of reduction, analyzing their effectiveness across various circuit classifications. Furthermore, we explore the effect of varying the partition size in the Lookahead-based brute-force algorithm, observing that increasing the partition size consistently decreases the number of non-reducible cases. 


\textbf{Paper Structure:}
Section~\ref{sec:methods} outlines the optimization strategies employed, including circuit generation for experimentation and the approximation techniques applied. Section~\ref{sec:evaluation} presents a comparative analysis of these algorithms, while Section~\ref{sec:conclusion} draws conclusions.

\section{Optimization Strategies} \label{sec:methods}

\subsection{Generating Diverse Circuits}

Our dataset spans 2,250 quantum circuits by varying qubits (10–100, 10 values), columns (15 values, 1 to $10 \times$ qubits), and T gates (15 values, $\max(\text{qubits}, \text{columns})$ to qubits $\times$ columns). Scaling columns up to $100 \times$ qubits ensures coverage of both shallow, high-parallelism and deep, high-depth circuits.
A random circuit generator simulates quantum circuits as a 2D array, where rows represent qubits and columns define depth. Each circuit is parameterized by qubits, columns, and T gates, which are randomly placed while ensuring each qubit and column contains at least one. Columns are assigned a random phase (`$+$' or `$-$'), and all gates initially default to identity (`I'). The remaining T gates are uniformly distributed, preserving a balanced structure.
We classify circuits using a three-layer framework based on depth, T-gate density, and qubit system size, yielding 27 categories ($3 \times 3 \times 3$). Depth is determined by column count percentiles: Shallow (S), Medium (M), and Deep (D). T-gate density, defined as $T_{\text{Gate Density}} = \frac{\text{Total T Gates}}{\text{Qubits} \times \text{Columns}}$, is categorized as Low (L), Medium (M), or High (H). Qubit system size is classified as Small (S), Medium (M), or Large (L). Circuits are labeled DTQ (Depth-T-Gate-Qubit), e.g., S-L-S for \underline{S}hallow, \underline{L}ow T-Gate Density, \underline{S}mall Qubit system, or D-H-L for \underline{D}eep, \underline{H}igh T-Gate Density, \underline{L}arge Qubit system.

\subsection{Approximation Techniques}




\textbf{Greedy Approach:}
In Algorithm~\ref{alg:greedy}, columns are reduced sequentially by combining with adjacent columns. This operation is repeated until no further reduction is possible.
Although easy to implement, it processes all column pairs without reordering, which can lead to inefficiency, especially for larger circuits.

\begin{algorithm}
\fontsize{8.5pt}{9.5pt}\selectfont
\caption{Greedy Algorithm}
\label{alg:greedy}
\begin{algorithmic}[1]
\State \textbf{Input:} Circuit $C$ with $n$ qubits and $c$ columns
\State \textbf{Output:} Reduced circuit $C'$
\While{multiple columns exist}
    \For{each adjacent column pair $(C_i, C_{i+1})$}
        \If{$C_i$ and $C_{i+1}$ commute}
            \State Compute element-wise $C_{new} = C_i \cdot C_{i+1}$
            \If{$C_{new}$ is phase consistent}
                \State Replace $C_i$ with $C_{new}$
                \State Remove $C_{i+1}$
            \EndIf
        \EndIf
    \EndFor
\EndWhile
\State \textbf{return} Reduced circuit $C'$
\end{algorithmic}
\end{algorithm}

\textbf{Divide and Conquer (D\&C) Approach:}
Algorithm~\ref{alg:dnc} splits the columns into two halves recursively, reducing each half independently before merging the results.
This approach ensures that only necessary operations are performed, as smaller subproblems are solved first and then combined.
This algorithm can balance efficiency with simplicity while maintaining a clear logical structure. It determines the sequence in which columns are merged, but it does not reorder the columns themselves in any way.

\begin{algorithm}
\fontsize{8.5pt}{9.5pt}\selectfont
\caption{Divide and Conquer Algorithm}
\label{alg:dnc}
\begin{algorithmic}[1]
\State \textbf{Input:} Circuit $C$ with $n$ qubits and $c$ columns
\State \textbf{Output:} Reduced circuit $C'$
\Function{Reduce}{$start, end$}
    \If{$start = end$}
        \State \textbf{return} $\{ C[start] \}$ 
    \EndIf
    \State $mid \gets \lfloor (start + end)/2 \rfloor$
    \State $L \gets$ \Call{Reduce}{$start, mid$}
    \State $R \gets$ \Call{Reduce}{$mid+1, end$}

    \If{$L$ and $R$ commute}        
        \State $C_{new} \gets$ element-wise $L \cdot R$
        \If{$C_{new}$ is phase consistent}
            \State \textbf{return} $\{C_{new}\}$ 
        \EndIf
    \EndIf
    \State \textbf{return} $L \cup R$ 
\EndFunction
\State \textbf{return} \Call{Reduce}{$0, c-1$}
\end{algorithmic}
\end{algorithm}

\textbf{Graph-Based Approach:}
Algorithm~\ref{alg:graph} models the circuit as a graph where each column is represented as a node, and edges indicate the similarity between adjacent columns.
The weight of each edge is determined based on how many gates remain unchanged between two adjacent columns. A Minimum Spanning Tree (MST) is then computed to determine the optimal merging sequence.
The algorithm iterates over the sorted MST edges, merging columns based on their similarity, thereby minimizing redundant operations and prioritizing beneficial merges.
This approach is particularly useful when the order of merging columns significantly impacts performance. By leveraging an optimized sequence derived from graph traversal, this approach efficiently reduces the circuit by strategically reordering the columns.

\begin{algorithm}
\fontsize{8.5pt}{9.5pt}\selectfont
\caption{Graph-Based Algorithm}
\label{alg:graph}
\begin{algorithmic}[1]
\State \textbf{Input:} Circuit $C$ with $n$ qubits and $c$ columns
\State \textbf{Output:} Reduced circuit $C'$
\State Construct graph $G$ where nodes represent columns
\For{each adjacent column pair $(C_i, C_{i+1})$}
    \State $w_{i, i+1} \gets |C_i \cap C_{i+1}|$
    \State $G \gets G \cup \{(C_i, C_{i+1}, w_{i, i+1})\}$
\EndFor
\State $MST(G) \gets Minimum\_Spanning\_Tree(G)$
\State $MST \gets Sort(MST, \text{descending by } w_{i,j})$
\While{$|C| > 1$ \textbf{ and } $\text{MST} \neq \emptyset$}
    \State Select edge $(i, j)$ with highest weight    
        \If{$C_i$ and $C_j$ commute}
         \State Compute element-wise $C_{new} = C_i \cdot C_{i+1}$
        
            \If{$C_{new}$ is phase consistent}
                \State Replace $C_i$ with $C_{new}$
                \State Remove $C_j$
            \EndIf
        \EndIf    
\EndWhile
\State \textbf{return} Reduced circuit $C'$ 
\end{algorithmic}
\end{algorithm}

\textbf{Lookahead-Based Brute-Force Approach:}
Since the original brute-force approach as described in Section~\ref{sec:introduction} \emph{(Motivation)} examines all possible reorderings of columns to find the most optimal merging sequence, it is an NP-hard problem and becomes computationally infeasible for large circuits. To address this, we adopt a lookahead-based strategy, as shown in Algorithm~\ref{alg:bf_lookahead}, to reduce the number of columns while maintaining correctness efficiently. 
%
The algorithm iterates over the circuit, selecting consecutive groups of $k$ columns and applying brute-force reduction. The original subset is retained if the reduction does not yield a smaller or optimized result. Once all subsets are processed, the reduced circuit replaces the previous one, and the process repeats until no further reduction is possible. Since $k \neq c$, this approach is less powerful than the brute-force method but still optimizes column ordering within its partition size to find the most effective merging sequence. This iterative refinement efficiently minimizes the number of columns without exhaustively searching the entire space. We primarily use a partition size of $k = 4$. If we consider $k$ explicitly as a non-trivial value the time complexity becomes: $O(c \cdot k \cdot k! \cdot n \log c)$.

\begin{algorithm}
\fontsize{8.5pt}{9.5pt}\selectfont
\caption{Lookahead-Based Brute-Force Algorithm}
\label{alg:bf_lookahead}
\begin{algorithmic}[1]
    \State \textbf{Input:} Circuit $C$ with $n$ qubits and $c$ columns
    \State \textbf{Output:} Reduced circuit $C'$
    \State $k \gets partition\_size$ \Comment{(Default: $k \gets partition\_size \gets 4$)}
    \State $num\_columns \gets c$

    \While{$num\_columns > k$}
        \State $C' \gets \emptyset$
        \For{each $subset$ of $k$ columns in $C$}
            \State $subset_{reduced} \gets brute\_force\_algorithm(subset)$

            
            \State $C' \gets C' \cup subset_{reduced}$ 
        \EndFor

        \If{$| C' | = | C |$}
            \State \textbf{break} 
        \EndIf

        \State $C \gets C'$
        \State $num\_columns \gets | C |$
    \EndWhile

    \State \textbf{return} Reduced circuit $C'$
\end{algorithmic}
\end{algorithm}

\textbf{Comparative Summary:}
Table~\ref{tab:algorithm_complexities} summarizes the time and space complexity of various approaches for T-depth reduction in quantum circuits with $n$ qubits and $c$ columns. The brute-force approach has the highest time complexity due to its factorial dependence on the number of columns, making it impractical for large circuits. In contrast, the greedy approach significantly reduces complexity to $O(c^2 n)$ but remains inefficient as it processes column pairs sequentially. The divide-and-conquer, BF lookahead ($k=4$), and graph-based methods all achieve a lower time complexity of $O(c \log c \cdot n)$, demonstrating their advantage in scalability. 
In terms of space complexity, all methods except divide-and-conquer require $O(nc)$ space, which scales linearly with the number of qubits and columns. The divide-and-conquer approach requires additional space, $O(nc + \log c)$, due to recursive function calls. 

\begin{table}[]
\centering
\fontsize{8.5pt}{9.5pt}\selectfont
\caption{Time and Space Complexities of the Algorithms}
\begin{tabular}{l||cc}
\multicolumn{1}{c||}{\textbf{Algorithm}}             & \textbf{Time Complexity}  & \textbf{Space Complexity} \\ \hline \hline
\textbf{Brute - Force}                             & $O(c! \cdot nc^2)$ & $O(nc)$                   \\
\textbf{Greedy}                                    & $O(nc^2)$                & $O(nc)$                   \\
\textbf{Divide \& Conquer}                         & $O(nc \log c)$     & $O(nc + \log c)$          \\
\textbf{Graph - based}                             & $O(nc \log c)$     & $O(nc)$                   \\
\textbf{Lookahead \emph{(k = 4)}} & $O(nc \log c)$     & $O(nc)$                   \\ \hline \hline
\end{tabular}
\label{tab:algorithm_complexities}
\end{table}
\section{Comparison and Evaluation} \label{sec:evaluation}

\textbf{Comparing the Optimization Methods:}  
Out of the 2,250 circuits in our dataset, 1,485 (\(66\%\)) are reducible, meaning at least one algorithm achieves a reduction in the number of columns. Conversely, for the remaining 765 circuits, no algorithm provides any reduction, resulting in a \(0\%\) improvement. Among the reducible circuits, Table~\ref{tab:algorithm_performance} presents the number of cases where each approximation technique performs best, along with their average percentage reduction. Notably, BF Lookahead with \(K = 4\) performs best in \(90\%\) of cases, achieving an average reduction of \(51.53\%\).

\begin{table}[]
\centering
\fontsize{8.5pt}{9.5pt}\selectfont
\caption{Performance Comparison of Approximation Techniques}
\begin{tabular}{l||ccc}
\multicolumn{1}{c||}{\textbf{Algorithm}}            & \textbf{\# Cases} & \textbf{\% Cases} & \textbf{ Avg. \% Reduction} \\ \hline \hline
\textbf{Greedy}                                    & 53                & 3.569             & 3.80                  \\
\textbf{Divide \& Conquer}                         & 58                & 3.905             & 21.76                 \\
\textbf{Graph - based}                             & 37                & 2.491             & 27.81                 \\
\textbf{Lookahead \emph{(k = 4)}} & 1337              & 90.033            & 51.53                 \\ \hline \hline
\end{tabular}
\label{tab:algorithm_performance}
\end{table}

\begin{figure}
    \centering
    \includegraphics[width=1\linewidth]{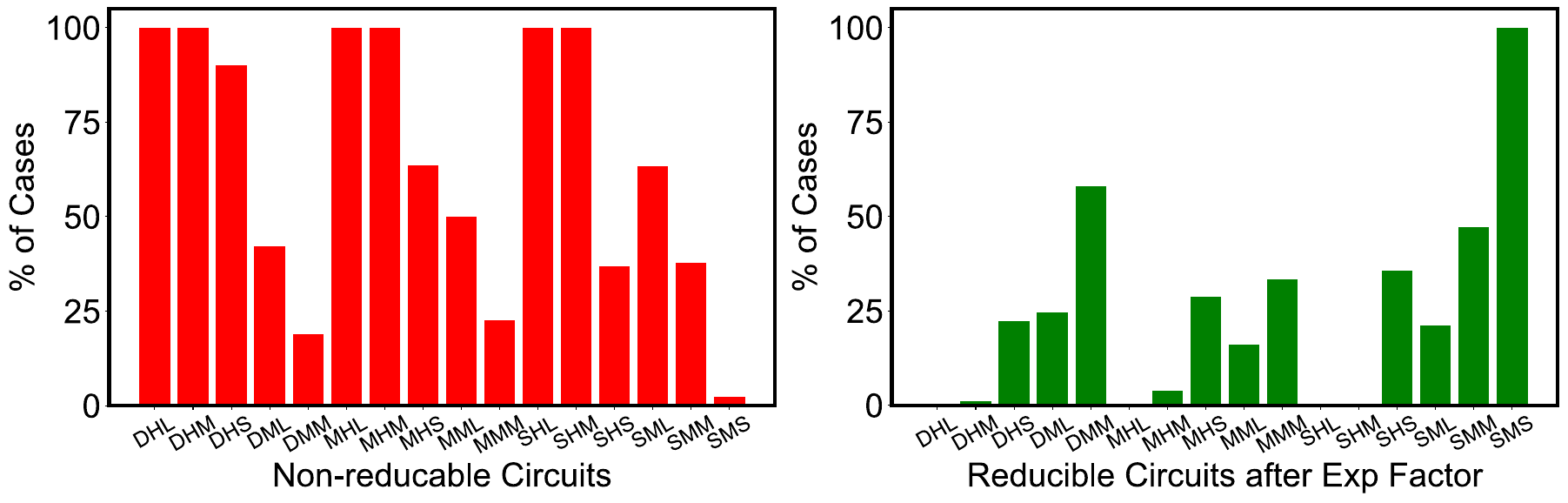}
    \caption{Classification of Non-Reducible and Newly Reducible Circuits: 
    (Left) Non-reducible circuit classifications with their percentage. (Right) Initially non-reducible classes, now showing the percentage successfully reduced using the expansion factor method.
    }
    \label{fig:reduce_non_reduce_classification}
    \vspace{-5mm}
\end{figure}

\textbf{Analysis of Reducibility Across Circuit Classifications:}
To determine which types of circuits are more amenable to reduction, we compare the classification of all circuits with those that remained unreduced across all algorithms (Fig. ~\ref{fig:reduce_non_reduce_classification} (left)). 
Circuits with a low T-gate density exhibit the highest likelihood of successful reduction. Among these, only a few classes such as DLM and MLL, contained unreduced instances, suggesting that additional factors, such as depth and qubit count, influence reduction feasibility. In contrast, circuits with medium T-gate density display a more varied response to optimization techniques. While some classes, such as SMS, were nearly fully reduced with only a small fraction remaining, others, including DMM, MMM, MML, and SMM, retained a nontrivial portion of unreduced circuits. This inconsistency suggests that while medium T-gate density does not inherently prevent reduction, its interaction with other structural properties like depth and qubit count, plays a significant role in determining optimization success.

In contrast, circuits with a high T-gate density (such as DHL, MHS, DHM, MHM, MHL, SHM, and SHL) consistently exhibit strong resistance to reduction.  
The complete failure to reduce certain classes underscores the significant computational complexity introduced by a high density of T gates, reinforcing the notion that T-gate placement is one of the most critical barriers to efficient circuit compression.

Although T-gate density exerts the strongest influence on reduction outcomes, circuit depth further compounds the difficulty of optimization. Deep circuits such as DHL, DHM, and MHL consistently show greater resistance to reduction, particularly when combined with a high T-gate density. 
Even when the reduction is partially successful, circuits such as DML and DHS retained a substantial fraction of unreduced instances, reinforcing the observation that depth significantly impacts reducibility. In contrast, shallow circuits demonstrate greater flexibility in reduction, as most shallow-depth classifications with low or medium T-gate density were successfully optimized. However, shallow circuits with high T-gate density, such as SHM, SHL, and SML, still exhibited resistance to reduction, indicating that high T-gate density can override the advantages typically associated with lower-depth circuits.

Qubit count also plays a role in reduction feasibility, though its influence is secondary to that of T-gate density and depth. Circuits with a large qubit count such as MHL, SHL, and DHL show a greater likelihood of reduction failure, particularly when combined with a high T-gate density. 
While small-qubit circuits generally showed better reduction performance, the presence of high T-gate density still restricted optimization regardless of qubit count. This suggests that while qubit count influences reduction difficulty, it does not impose as severe a constraint as T-gate density or circuit depth.

\emph{Summary of analysis:} The overall trends in reduction feasibility indicate that high T-gate density is the primary limiting factor in circuit optimization. 
Circuit depth amplifies this difficulty
especially when combined with a high density of T gates. Qubit count further contributes to reduction challenges, but its impact is less pronounced compared to T-gate density and depth. 

\textbf{Integration of the Expansion Factor into Circuits:}
The expansion factor alters the structure of a quantum circuit by increasing the number of columns and redistributing T gates through the insertion of redundant identity gates. This process effectively stretches the circuit while preserving its phase properties and logical integrity. 
Since high T-gate density is the primary limiting factor in circuit optimization, introducing an expansion factor can be beneficial despite increasing circuit depth proportionally. While this added depth raises the complexity of approximation approaches, the expansion technique still improves performance by inserting redundant identity gates, effectively reducing the overall T-gate density of the circuit.
The application of expansion depends on the relationship between the expansion factor and the number of qubits in the original circuit. In our experiments, we explore expansion factors ranging from 2 to 25.
If the number of qubits is perfectly divisible by the expansion factor, the circuit is expanded by simply repeating each column while splitting the qubits evenly. Each new column is assigned a portion of the original qubits, ensuring that the structure of the circuit remains intact. 
When the expansion factor is greater than the number of qubits, additional padding qubits are introduced (temporarily) before the expansion process. These extra qubits are initialized with identity gates that alternate between positive and negative phases to maintain symmetry. Once the required number of qubits is achieved, the circuit is expanded as in the previous case. After expansion, any extra qubits that were temporarily added are removed to ensure the final circuit retains its intended dimensions.
If the expansion factor is smaller than the number of qubits and the qubits cannot be evenly divided, the qubits are distributed dynamically across the expanded columns. The circuit ensures that the extra qubits are assigned fairly among the new columns while maintaining phase consistency. 
The expansion process allows for scalable circuit modifications without altering the logical behavior of the computation. An example of an original circuit consisting of four columns, four qubits, and sixteen T gates is illustrated in Fig.~\ref{fig:times_factor_figure}, demonstrating expansion factors of 2 and 4. 

\textbf{Analysis of the Expansion on Previously Unreduced Circuits:}
We expand the circuits that remained unreduced and optimize them using the previously mentioned approximation techniques. 
Our analysis reveals that 15\% of these circuits were successfully reduced through expansion.
Fig.~\ref{fig:reduce_non_reduce_classification} (right) illustrates the percentage of circuit classes (with respect to previously irreducible) that are successfully reduced.
The results indicate varying degrees of success across different circuit classes. Some classes saw complete elimination of previously unreduced circuits, while others showed only marginal or no reduction at all. This analysis evaluates the effectiveness of the expansion across different depth, T-gate density, and qubit count configurations. The expansion was highly effective for SMS, completely reducing all previously unreduced circuits. It also provided significant improvement for DMM and SMM, indicating that medium-depth circuits with medium T-gate density benefit from expansion. However, for circuits with high T-gate density, especially in deep architectures, the method was largely ineffective, as evidenced by classes such as DHL, MHL, SHM, and SHL, which showed no reduction. The low success rate for deep and high-T circuits suggests that their optimization constraints are more fundamental, likely requiring additional techniques such as more advanced gate synthesis methods or alternative decompositions. These findings highlight that while expansion enhances reduction in many cases, it is not universally effective. The primary limiting factor remains the T-gate density, with high-T circuits showing the strongest resistance. 

\begin{figure}
    \centering
    \includegraphics[width=1\linewidth]{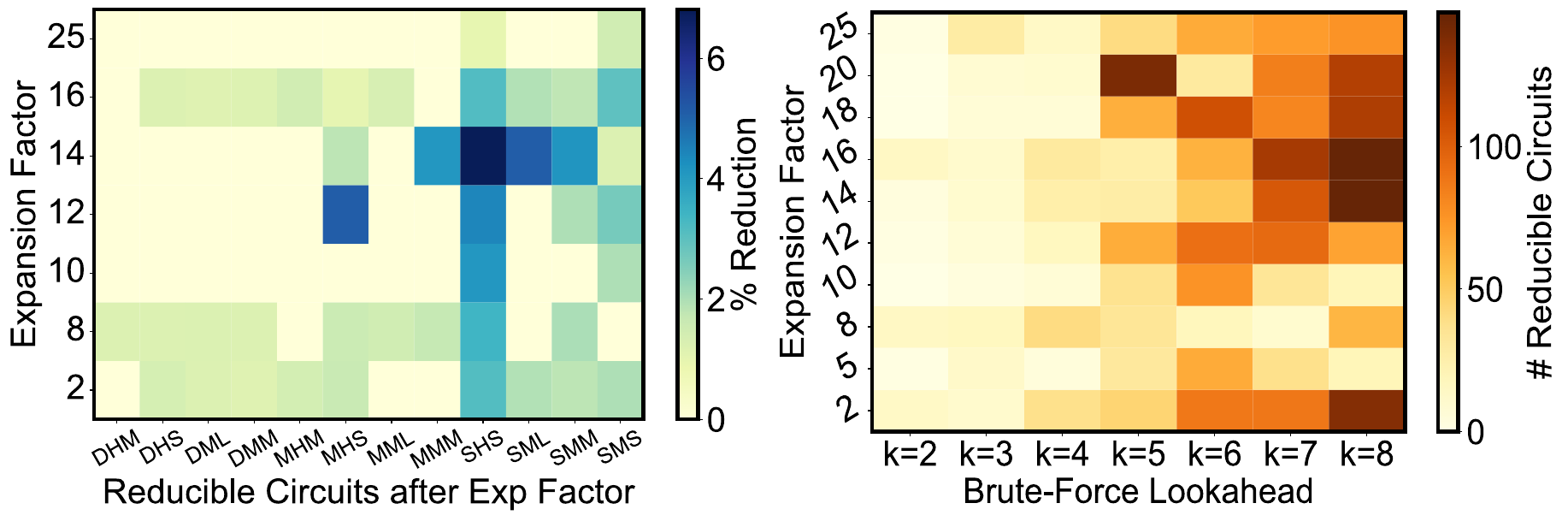}
    \caption{Impact of Expansion Factor and Lookahead Partition on Circuit Reduction: 
    (Left) Reducible circuit classes vs. expansion factor, showing average column reduction with no strong correlation. (Right) Reducible circuit count across expansion factors and partition sizes, where increasing partition size helps, but no clear trend emerges for expansion factor.
    }
    \label{fig:healtmaps}
\end{figure}

We observe that among all circuits reducible using the expansion, 98\% were successfully reduced using the Brute-Force Lookahead algorithm with a partition size of 4. Given this overwhelming effectiveness, we now focus exclusively on the Brute-Force Lookahead algorithm for further analysis.
%
Fig.~\ref{fig:healtmaps} (left) illustrates the classes of circuits that become reducible plotted against the expansion factor values. For each case, we present the average percentage reduction in the number of columns. While a strong correlation between the expansion factor and the average percentage reduction is not immediately apparent, expansion factors in the range of 12 to 16 appear to be the most effective in most cases.

\begin{table}[]
\centering
\fontsize{8.5pt}{9.5pt}\selectfont
\caption{Effect of Partition Size on Success Rate and Performance}
\begin{tabular}{c||ccccccc}
\textbf{Partition (k)}  & \textbf{2}               & \textbf{3}               & \textbf{4}               & \textbf{5}               & \textbf{6}               & \textbf{7}               & \textbf{8} \\ \hline \hline
\textbf{\% Succ. Cases} & \multicolumn{1}{l}{8.17} & \multicolumn{1}{l}{6.59} & \multicolumn{1}{l}{14.3} & \multicolumn{1}{l}{22.2} & \multicolumn{1}{l}{24.2} & \multicolumn{1}{l}{24.5} & 24.8       \\
\textbf{Avg. \% Redn.}  & 8.91                     & 8.36                     & 9.52                     & 8.33                     & 8.52                     & 10.7                     & 11.8       \\ \hline \hline
\end{tabular}
\label{tab:lookahead_performance_k}
\end{table}

\begin{figure}
    \centering
    \includegraphics[width=1\linewidth]{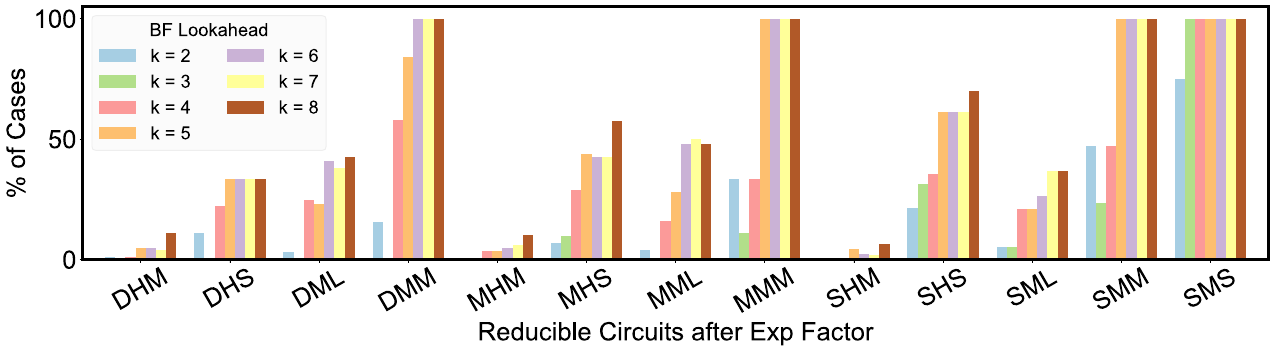}
    \caption{Effect of Partition Size on Circuit Reducibility: 
    Percentage of initially unreducible circuits that become reducible when applying different partition sizes in the brute-force lookahead algorithm. Increasing the partition size consistently improves reducibility.
    }
    \label{fig:all_k_bar_plot}
\end{figure}

Since our primary objective is to maximize the number of reducible circuits, we explored this further by varying the lookahead partition size $k$ for the brute-force lookahead approach in Algorithm~\ref{alg:bf_lookahead} to determine whether adjusting this parameter increases the number of reducible cases, thereby reducing the count of non-reducible circuits. 
Starting with an initial set of 765 non-reducible circuits, we observe that as the lookahead partition size increases, more circuits become reducible (Fig. ~\ref{fig:healtmaps} (right)). However, no clear trend emerges regarding the direct impact of the expansion factor on this behavior. 
Table~\ref{tab:lookahead_performance_k} shows the overall percentage of successful cases where non-reducible circuits become reducible, along with the trend in average reduction percentage as partition size increases. The overall success rate improves with larger partitions but stabilizes around  $k = 6$ , suggesting that further increasing the partition size has little additional impact. Meanwhile, the average reduction percentage remains largely consistent across partition sizes, with only slight improvements.
Fig.~\ref{fig:all_k_bar_plot} further breaks this down by circuit category, confirming that while increased partition size enhances reducibility across all types, the gains plateau consistently around $k = 6$.

\section{Conclusion} \label{sec:conclusion}

In this work, we addressed the NP-hard problem of T-depth reduction in quantum circuits, a crucial factor in optimizing resource efficiency for fault-tolerant quantum computing. We explored multiple approximation techniques and introduced an expansion factor-based identity gate insertion strategy to enhance circuit reducibility. Additionally, we examined the impact of expansion factors and partition size variations on the effectiveness of T-depth reduction.
These insights contribute to a deeper understanding of circuit reduction strategies and their scalability, ultimately aiding in the minimization of magic state overhead in large-scale quantum architectures. 

\section*{Acknowledgment}

The work is supported in parts by the National Science Foundation (NSF) (CNS-1722557, CCF-1718474) and gifts from Intel.

\bibliographystyle{unsrt}
\bibliography{refs}

\begin{thebibliography}{10}

\bibitem{preskill1998reliable}
John Preskill.
\newblock Reliable quantum computers.
\newblock {\em Proceedings of the Royal Society of London. Series A: Mathematical, Physical and Engineering Sciences}, 454(1969):385--410, 1998.

\bibitem{lidar2013quantum}
Daniel~A Lidar et~al.
\newblock {\em Quantum error correction}.
\newblock Cambridge university press, 2013.

\bibitem{roffe2019quantum}
Joschka Roffe.
\newblock Quantum error correction: an introductory guide.
\newblock {\em Contemporary Physics}, 60(3):226--245, 2019.

\bibitem{clerk2010introduction}
Aashish~A Clerk et~al.
\newblock Introduction to quantum noise, measurement, and amplification.
\newblock {\em Reviews of Modern Physics}, 82(2):1155--1208, 2010.

\bibitem{fowler2012surface}
Austin~G Fowler et~al.
\newblock Surface codes: Towards practical large-scale quantum computation.
\newblock {\em Physical Review A—Atomic, Molecular, and Optical Physics}, 86(3):032324, 2012.

\bibitem{kitaev2003fault}
A~Yu Kitaev.
\newblock Fault-tolerant quantum computation by anyons.
\newblock {\em Annals of physics}, 303(1):2--30, 2003.

\bibitem{google2023suppressing}
Suppressing quantum errors by scaling a surface code logical qubit.
\newblock {\em Nature}, 614(7949):676--681, 2023.

\bibitem{bravyi2012magic}
Sergey Bravyi et~al.
\newblock Magic-state distillation with low overhead.
\newblock {\em Physical Review A—Atomic, Molecular, and Optical Physics}, 86(5):052329, 2012.

\bibitem{litinski2019magic}
Daniel Litinski.
\newblock Magic state distillation: Not as costly as you think.
\newblock {\em Quantum}, 3:205, 2019.

\bibitem{kliuchnikov2012fast}
Vadym Kliuchnikov et~al.
\newblock Fast and efficient exact synthesis of single qubit unitaries generated by clifford and t gates.
\newblock {\em arXiv preprint arXiv:1206.5236}, 2012.

\bibitem{brown2017poking}
Benjamin~J Brown et~al.
\newblock Poking holes and cutting corners to achieve clifford gates with the surface code.
\newblock {\em Physical Review X}, 7(2):021029, 2017.

\bibitem{haah2018codes}
Jeongwan Haah et~al.
\newblock Codes and protocols for distilling $ t $, controlled-$ s $, and toffoli gates.
\newblock {\em Quantum}, 2:71, 2018.

\bibitem{bravyi2005universal}
Sergey Bravyi et~al.
\newblock Universal quantum computation with ideal clifford gates and noisy ancillas.
\newblock {\em Physical Review A—Atomic, Molecular, and Optical Physics}, 71(2):022316, 2005.

\bibitem{litinski2018quantum}
Daniel Litinski et~al.
\newblock Quantum computing with majorana fermion codes.
\newblock {\em Physical Review B}, 97(20):205404, 2018.

\bibitem{gottesman1998heisenberg}
Daniel Gottesman.
\newblock The heisenberg representation of quantum computers.
\newblock {\em arXiv preprint quant-ph/9807006}, 1998.

\bibitem{litinski2019game}
Daniel Litinski.
\newblock A game of surface codes: Large-scale quantum computing with lattice surgery.
\newblock {\em Quantum}, 3:128, 2019.

\bibitem{van2023optimising}
John van~de Wetering et~al.
\newblock Optimising quantum circuits is generally hard.
\newblock {\em arXiv preprint arXiv:2310.05958}, 2023.

\end{thebibliography}

\end{document}